\documentclass[10pt,a4paper]{article}
\usepackage{amsmath}
\usepackage{amsfonts}
\usepackage{amssymb}
\usepackage{graphicx}
\usepackage{textcomp}
\usepackage{url}
\usepackage{lipsum}
\usepackage{setspace}
\usepackage{color}
\usepackage[numbers,sort&compress]{natbib}
\usepackage[usenames,dvipsnames,svgnames,table]{xcolor}
\usepackage[pdfstartview=FitH]{hyperref}
\usepackage{hyperref}
\newsavebox{\bigleftbox}
\hypersetup{
 colorlinks,
 citecolor=Blue,
 linkcolor=Blue,
 urlcolor=Blue
}

\makeatletter
 \def\footnoterule{\kern-3\p@
   \noindent\hrulefill \kern 2.8\p@} % the \hrule is .4pt high
\makeatother
\usepackage[left=3.00cm, right=3.00cm, top=2.00cm, bottom=2.00cm]{geometry}

\usepackage[normalem]{ulem} % pra usar \sout entre nós apenas
\usepackage{fancyhdr}

\title{\textbf{Elastocaloric Effect in Graphene Kirigami}}
\author{ 
	Luiz A. Ribeiro Junior$^{1,\dag}$,
        Marcelo L. Pereira Junior$^{2,\ddag}$, and
        Alexandre F. Fonseca$^{3,\S}$
	}
\date{}

\begin{document}
    \maketitle
	\vspace{-0.6cm}
	\begin{center}\small
		$^1$\textit{University of Brasília, Institute of Physics, Brasília, Brazil}\\
		$^2$\textit{University of Bras\'{i}lia, Faculty of Technology, Department of Electrical Engineering, Bras\'{i}lia, Brazil}\\
		$^3$\textit{Applied Physics Department, Gleb Wataghin Institute of Physics, University of Campinas, Campinas, S\~ao Paulo, Brazil.}\\ \vspace{0.2cm} 
		\phantom{.}\hfill
		$^{\dag}$\url{ribeirojr@unb.br}\hfill
		$^{\ddag}$\url{marcelo.lopes@unb.br}\hfill
		$^{\S}$\url{afonseca@ifi.unicamp.br}\hfill
		\phantom{.}
	\end{center}
	
% ---------------------- %
%\blfootnote{}
%\blfootnote{}

\onehalfspace

\noindent\textbf{Abstract:} Kirigami, a traditional Japanese art of paper-cutting, has recently been explored for its elastocaloric effect (ECE) in kirigami-based materials (KMs), where applying strain induces temperature changes. In this study, we investigate the ECE in a nanoscale graphene kirigami (GK) monolayer, representing the thinnest possible KM, to better understand this phenomenon. Through molecular dynamics simulations, we analyze the temperature change and coefficient of performance (COP) of the nanoscale GK architecture. Our findings reveal that while GKs lack the intricate temperature changes observed in macroscopic KMs, they exhibit a substantial temperature change of approximately 9.32 K (23 times higher than that of macroscopic KMs, which is about 0.4K) for heating and -3.50 K for cooling. Additionally, they demonstrate reasonable COP values of approximately 1.57 and 0.62, respectively. It is noteworthy that the one-atom-thick graphene configuration prevents the occurrence of the complex temperature distribution observed in macroscopic KMs.

%Kirigami is a traditional Japanese art of paper-cutting that has become an essential tool for creating structures with adjustable and programmable mechanical properties. Recently, researchers measured the elastocaloric effect (ECE) in kirigami-based materials (KMs), which refers to the temperature change induced by applying strain. While KMs offer high flexibility and can achieve large strains with minimal stress, the observed temperature changes resulting from ECE were relatively low, reaching a maximum of approximately 0.4 K. Interestingly, KMs demonstrated complex temperature variations, with simultaneous increases and decreases in different regions when the external strain was applied. To delve deeper into this phenomenon, we investigated the ECE in a graphene kirigami (GK) monolayer at the nanoscale, representing the thinnest possible KM. Through fully-atomistic molecular dynamics simulations, we analyzed the change in temperature and coefficient of performance (COP) of the nanoscale GK architecture. COP, the ratio of the heat removed from the cold region to the total work performed on the system per thermodynamic cycle, was also evaluated. Our results revealed that although GKs lack the intricate temperature changes observed in macroscopic KMs, they exhibit a relatively large temperature change of approximately 9.32 K for heating and -3.50 K for cooling, accompanied by a reasonable COP of approximately 3. Notably, the one-atom-thick configuration of graphene prevents the occurrence of the complex temperature distribution observed in macroscopic KMs.

\section{Introduction}

Graphene is a two-dimensional allotrope of carbon atoms arranged in a hexagonal lattice \cite{geim2009graphene}. It was first isolated in 2004 using the mechanical exfoliation technique \cite{novoselov2004electric}. Graphene's exceptional combination of mechanical \cite{papageorgiou2017mechanical}, electronic \cite{neto2009electronic}, thermal~\cite{TEGraphene2009NatNano,TCGraphene2014NatComm} and optical properties \cite{falkovsky2008optical} has made it one of the most extensively studied systems in materials science, with vast potential for use in various applications such as flat electronics \cite{lui2009ultraflat}, energy storage \cite{raccichini2015role}, and biomedical engineering \cite{chen2015two}. However, incorporating graphene into flexible devices poses challenges because of its intrinsically high stiffness, which makes it difficult to bend or stretch. In addition, graphene is brittle, with only a few percent tensile fracture strains \cite{akinwande2017review}.

Graphene Kirigami (GK) is a cutting-edge technique miming the ancient Japanese art of paper-cutting to manipulate graphene \cite{blees2015graphene}. By introducing intricate patterns and tailored structures, GK overcomes the low stretchability barrier of graphene, resulting in unique mechanical properties \cite{qi2014atomistic,hanakata2018accelerated}. Compared to a regular graphene sheet, GK nanostructures exhibit significantly improved reversible stretchability, achieved through multidimensional deformation capabilities enabled by the nanoscale Kirigami architectures \cite{hua2017large}. The GK's in-plane stretching and out-of-plane deformation allow for increased flexibility \cite{hua2017large}. With flexible devices being the mainstream trend in modern optoelectronics, Kirigami-like materials have become ideal candidates for developing novel applications.

The fabrication of GK structures has yielded remarkable success \cite{blees2015graphene,grosso2015bending,wei2016thermal,gao2022graphene,mortazavi2017thermal,cai2016effects,yong2020kirigami,ning2018assembly,lee2020multiaxially,xu2019highly}. Initial work by Blees and colleagues demonstrated the feasibility of GKs, showcasing robust microscale structures with adjustable mechanical properties in graphene \cite{blees2015graphene}. Another experimental study highlighted the strain-insensitive electrical properties of GK-based electrodes, enduring up to 240\% tensile strain and various strain states, including stretching, twisting, and shearing \cite{yong2020kirigami}. Furthermore, atomistic molecular dynamics (MD) simulations have extensively investigated GK's stretchability, fracture resistance, and permeability \cite{wei2016thermal,qi2014atomistic,hanakata2018accelerated,hua2017large,gao2022study,li2020anomalous,gao2022graphene,mortazavi2017thermal,gamil2020mechanical}. Notably, these studies unveiled a noteworthy trend: GK can enhance the fracture strains of graphene by approximately threefold compared to standard monolayer graphene. Consequently, the realm of Kirigami-inspired nanostructures presents exciting prospects for the fabrication of highly flexible and stretchable devices using graphene \cite{jang2016graphene} and other diverse 2D materials \cite{hanakata2016highly,wang2019nanoindentation,han2017super}.

A temperature change induced by external strain, known as the elastocaloric effect (ECE), has great potential for next-generation thermal management technologies due to its environmental friendliness and economic benefits~\cite{manosa2013,moya2014,moya2020}. 
To make this technology more efficient, efforts have been made to find highly efficient elastocaloric materials~\cite{chauhan2015,schmidt2016,jaka2016,cazorla2016,ray2019,muniz2020}. In particular, in view of the growing interest in miniaturized refrigeration~\cite{Imran2021}, at the nanoscale, the study of the ECE in nanostructures has already begun \cite{lisenkov2016elastocaloric,zhang2017elastocaloric,skavcej2018elastocaloric,CantuarioANNPHY2019,warzoha2020grain,LiSCICHINE2020,TatianaPRB2022,liu2022large,zhao2022room}. Recently, a new approach to modulate the elastocaloric temperature change using a Kirigami-inspired material was proposed \cite{hiraiADVFMATT2022}. Results showed that the ultrahigh stretchability of kirigami combined with its lateral flexibility allows for the generation of focused heat absorption and release simultaneously during the same application of tensile or compressive strain, resulting in a versatile local cooling/heating performance \cite{hiraiADVFMATT2022}. The method can be applied to any elastocaloric material and pave the way for designing versatile solid-state heat pumps for flexible electronics. So far, this phenomenon in Kirigami has not been studied on the atomic scale. Moreover, based on the successful synthesis of GKs, a study on the ECE in these materials may open new channels for developing novel thermal management technologies at the nanoscale.

In this study, we employed fully-atomistic MD simulations to investigate the ECE in a model GK lattice at the nanoscale. To simulate the ECE in GKs, we modeled a thermal management device under Otto-like thermodynamic cycles, including two adiabatic expansion/contraction and two isochoric heat exchange processes. Using our computational protocol, we calculate the GK temperature variation in absorbing and releasing heat and their coefficients of performance.  

\section{Methodology}

\begin{figure}[htb!]
\begin{center}
\includegraphics[width=\linewidth]{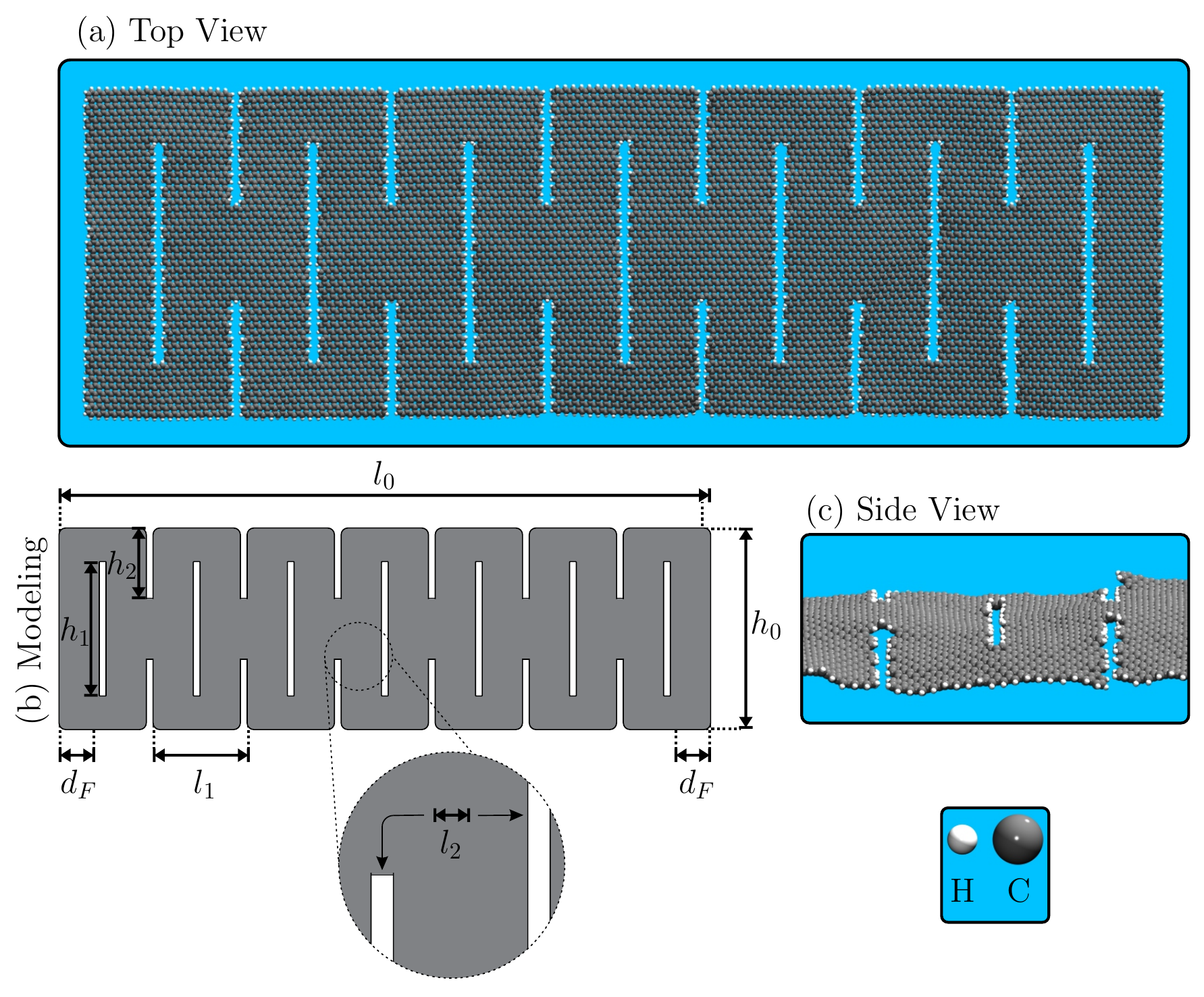}
\caption{Schematic representation of the model GK lattice studied here. Panels (a-c) show the GK top view, lattice parameters definitions, and KG side view. In the color scheme, the grey and white spheres represent the carbon and hydrogen atoms, respectively.}
\label{fig1}
\end{center}
\end{figure}

To investigate the ECE in a model GK lattice at the atomic scale, we used the LAMMPS \cite{thompson2022lammps} code with the Adaptive Intermolecular Reactive Empirical Bond Order (AIREBO) \cite{stuart2000reactive} potential in reactive MD simulations. This potential accurately addresses carbon-based nanostructures and provides reliable potential energy values to study their mechanical and thermal properties \cite{ruoffSCIENCE2010,NeekAmalPRB2011,GaoJMECHPHYSSOLIDS2014,fonsecaTECPRB2014,fonseca0TECJPCC2015,LeviJCCP2020}. Besides, the LAMMPS/AIREBO methodology has been successfully used to describe the ECE in carbon nanomaterials \cite{lisenkov2016elastocaloric,CantuarioANNPHY2019,LiSCICHINE2020,zhao2022room,TatianaPRB2022}. The GK model used in this study is shown in Figure \ref{fig1} and has the following properties: $l_0=332$~\AA, $l_1=48$~\AA, $l_2=4.8$~\AA, $h_0=100$~\AA, $h_1=68$~\AA, $h_2=35$~\AA, $d_f=19$~\AA, 12012 carbon atoms, and 1244 hydrogen atoms. 

\begin{figure}[htb!]
\begin{center}
\includegraphics[width=0.6\linewidth]{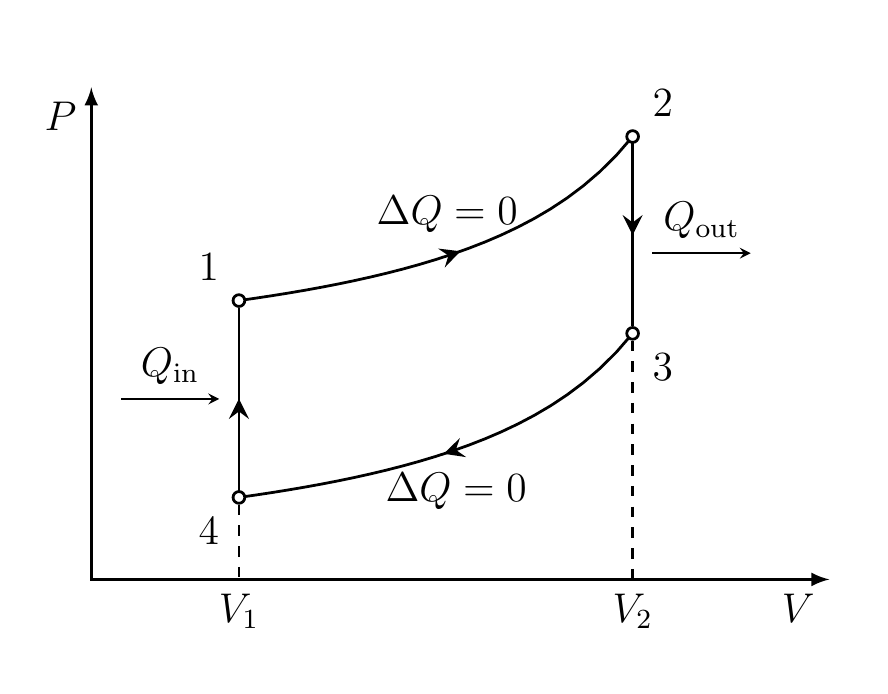}
\caption{Schematic representation of a P-V-like diagram that illustrates the thermodynamic cycle ($1\rightarrow 2\rightarrow 3\rightarrow 4\rightarrow 1$) used to study the ECE in GK. Processes $1\rightarrow2$ and $3\rightarrow4$ correspond to the application of adiabatic tensile strain and release, respectively, while processes $2\rightarrow3$ and $4\rightarrow1$ represent isochoric heat exchange with a reservoir at 300 K.}
\label{fig2}
\end{center}
\end{figure}

The GK structure was studied under a thermodynamic cycle similar to the Otto cycle, as shown in Figure \ref{fig2}. The cycle consists of two adiabatic expansion/release processes and two volume fixed heat exchange processes with a thermal reservoir at 300 K, in the order $1\rightarrow2\rightarrow3\rightarrow4\rightarrow1$. Processes $1\rightarrow2$ and $3\rightarrow4$ correspond to applying adiabatic tensile strain and release, %(compression), 
respectively. Processes $2\rightarrow3$ and $4\rightarrow1$ represent isochoric heat exchange with a reservoir at 300 K. During the stretching/compression process, the left edge of the kirigami remained fixed, while the right edge was stretched/compressed. In the isochoric heat exchange, both edges remained fixed.  

Initially, we minimized the GK structure using a force and relative energy tolerance of $1.0\times 10^{-10}$ eV/\r{A} and $1.0\times 10^{-10}$, respectively. The initial GK length ($l_0$, refer to Figure \ref{fig1}) is 337.20 \r{A}, and its equilibration was then conducted at 300 K, using a constant NPT ensemble integration at null pressure using the Nos\'e-Hoover barostat for 200 ps to ensure no remaining stress. The thermodynamic cycle was performed as follows: first, we carried out step ($1\rightarrow2$, expansion) without a thermostat, subjecting the system to tensile strain from 0\% to 30\% at a strain rate of $1\cdot10^{-4}$ fs$^{-1}$ during 32 ps, using a time step of 0.02 fs. Second, step ($2\rightarrow3$, isochoric heat exchange) involved a 200 ps thermal equilibration using the Langevin thermostat and NVE ensemble for a time step of 0.2 fs. Step ($3\rightarrow4$, release) was performed without a thermostat, causing the system to undergo tensile strain from 30\% to 0\% at a strain rate of $-1\cdot10^{-4}$ fs$^{-1}$ during 32 ps, using a time step of 0.02 fs again. Finally, step ($4\rightarrow1$, isochoric heat exchange) involved another 200 ps thermal equilibration using the Langevin thermostat and NVE ensemble for a time step of 0.2 fs.

The real ECE temperature variation ($\Delta T_\text{REAL}$) is obtained from correcting the results from classical MD simulations due to quantum mechanics effects on the specific heat ($c$) of graphene at 300 K \cite{lisenkov2016elastocaloric}: $\Delta T_\text{REAL}=(c_\text{MD}/c_\text{REAL})\Delta T_\text{MD}$. Here, ``REAL" and ``MD" refer to corrected and simulated values, respectively. The $(c_\text{MD}/c_\text{REAL})=2.53$ factor was obtained using $c_\text{REAL}=700$ $\mathrm{J}\cdot\mathrm{Kg}^{-1}\cdot\mathrm{K}^{-1}$ from Ref.~\cite{liNANOSCALE2017} and $c_\text{MD}=1770$ $\mathrm{J}\cdot\mathrm{Kg}^{-1}\cdot\mathrm{K}^{-1}$ from our calculations. $c_\text{MD}$ was derived from a series of equilibrium simulations for GK subjected to distinct temperature regimes, as discussed later. 

\section{Results}

\begin{figure}[htb!]
\begin{center}
\includegraphics[width=\linewidth]{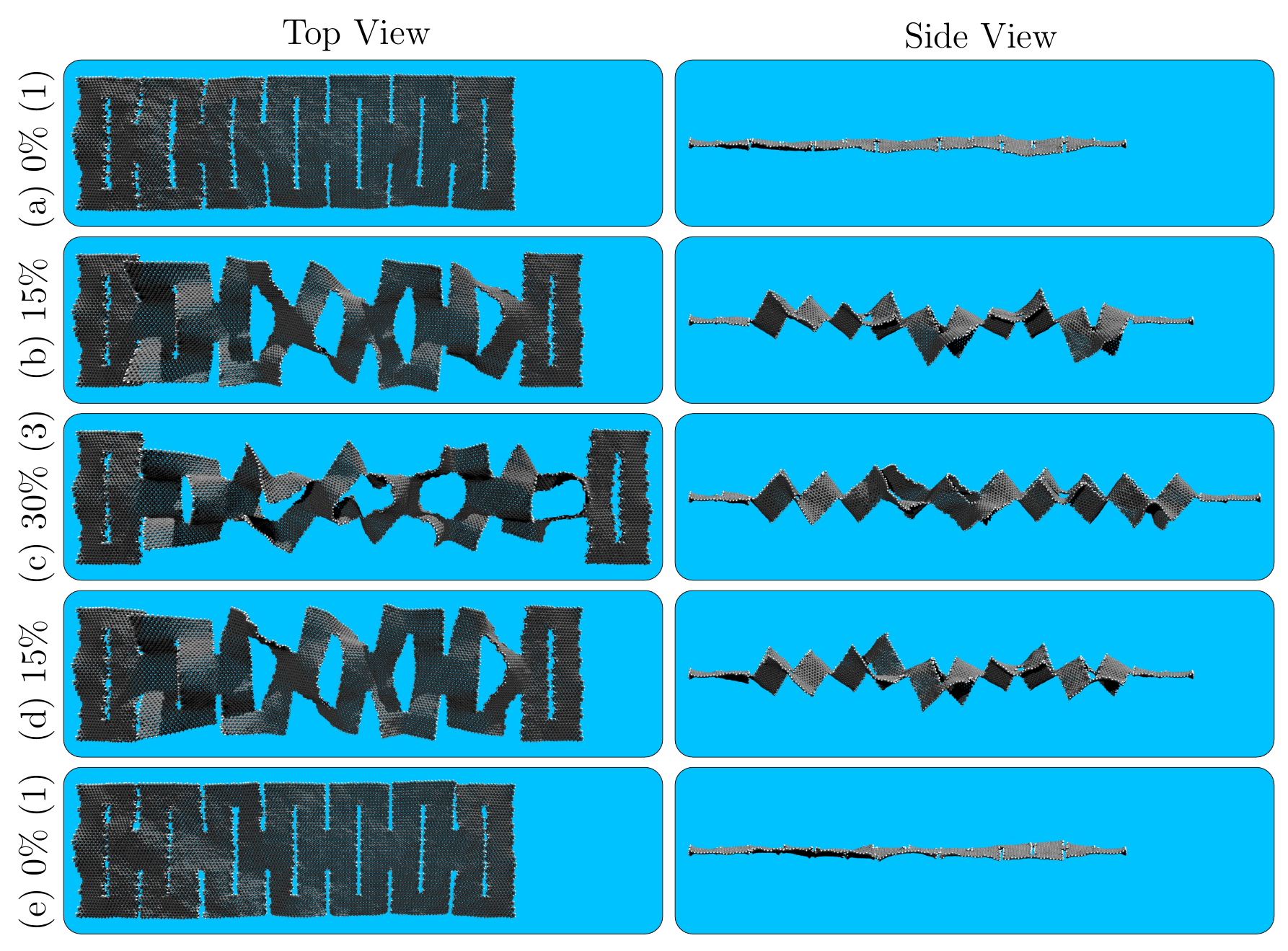}
\caption{Representative MD snapshots for the GK lattice subjected to the uniaxial strain in the $x$-direction at $300$ K. Panels (a-e) illustrate the top and side views for the GK structure at 0\% of strain before stretching, 15\% of stretching, 30\% of stretching, 15\% of releasing, and 0\% of strain after releasing, respectively. In the color scheme, the grey and white spheres represent the carbon and hydrogen atoms, respectively.}
\label{fig3}
\end{center}
\end{figure}

We begin our discussion by showing the representative MD snapshots illustrating GK's stretching/releasing process (see Figure \ref{fig3}). In Figure \ref{fig3}(a), we present the top and side views for the KG structure at 0\% of strain and equilibrated at 300 K. Even considering the atomic displacements imposed by the temperature effects, one can note that the GK is almost planar and its pores have the same size. It is worth mentioning that between 0\%-5\% of strain, the GK deformation occurs via C-C elastic bond stretching in the direction of the applied strain, similar to the stretching mechanism of graphene nanoribbons \cite{bu2009atomistic}. In this low-strain regime, neither out-of-plane flipping nor rotation of the GK pieces is observed.  

At a strain of 15\% (see Figure \ref{fig3}(b)), noticeable flipping and rotation occurred in certain sections of GK, resulting in varying magnitudes of out-of-plane deflections instead of the anticipated continuous bond stretching in uniformly strained sheets \cite{ni2010anisotropic}. Unlike pristine graphene nanoribbons that experience failure when reaching a critical strain \cite{pei2010molecular}, these deflections enable the GK monolayer to withstand significantly larger deformations without fracturing. 

It is crucial to contrast our findings with the study conducted by Hirai and coworkers \cite{hiraiADVFMATT2022}, who investigated the electromechanical cooling effect (EMCE) in various macroscopic kirigami materials (KMs). Using sophisticated lock-in thermography techniques, they measured temperature changes across the kirigami surface and discovered that cooling and warming regions coexist within different regions of the structure during the same type of straining (tensile or compressive). This phenomenon arises from the bending of hole edges within the kirigami plane under tensile strain, inducing both tensile and compressive strains in the outer and inner regions of the bent structure. They also demonstrated that the cooled and warmed regions precisely corresponded to the tensile-strained and compressed parts of the kirigami, respectively. However, we should point out that this effect does not (and cannot) occur in one-atom-thick kirigami. 

In our simulations, depicted in Figures \ref{fig3}(b-d), the deformation mechanism of GK involves two distinct regions: elastic stretching of C-C bonds (resulting in an ECE increase of temperature) and out-of-plane buckling. Unlike macroscopic kirigami, these buckled regions neither experienced cooling nor warming. This behavior diverges from the EMCE observed in macroscopic KMs. The reason behind this disparity is relatively straightforward. When a thin but macroscopic ribbon flexes, it induces tensile and compressive strains in its outer and inner sections. However, flexing a one-atom-thick nanoribbon, as demonstrated in previous research \cite{ZhangPRL2011}, does not lead to any bond strain. Therefore, in our case, the out-of-plane deflections did not generate strains in the C-C bonds. 

The accumulation of stress in GK is supported by the von Mises stress distribution calculated from an MD study of GK under tension \cite{hua2017large}. Consequently, as neither tensile nor compressive strains occur in the out-of-plane deflected regions of GK, there is no heat release or absorption in these areas. In contrast, the heated regions consist of parallel lines of stretched C-C bonds that remain within the GK plane \cite{hua2017large}. This stretching mechanism observed in GK opposes the findings of experiments \cite{hiraiADVFMATT2022}, where stretching elastocaloric materials with a macroscopic Kirigami structure resulted in simultaneous heat absorption and release due to non-uniform internal stress generated by uneven strain distribution along the cutting patterns.

Figure \ref{fig3}(c) shows that the highly stretchable GK pattern, which exhibits no bond breaking or fracture, is achieved at a critical strain of 30\%. In this configuration, each GK unit is connected by a small ribbon, allowing the monolayer to be almost foldable. Additionally, a uniform out-of-plane deflection pattern is observed. This configuration limits the interplay between the warming and cooling of GK via ECE temperature variation. Figures \ref{fig3}(d) and \ref{fig3}(e) illustrate the GK monolayer configuration during the tensile-releasing mechanism, %(compressive strain), 
which features are similar to those described for Figures \ref{fig3}(a) and \ref{fig3}(b), respectively.   

\begin{figure}[htb!]
\begin{center}
\includegraphics[width=0.48\linewidth]{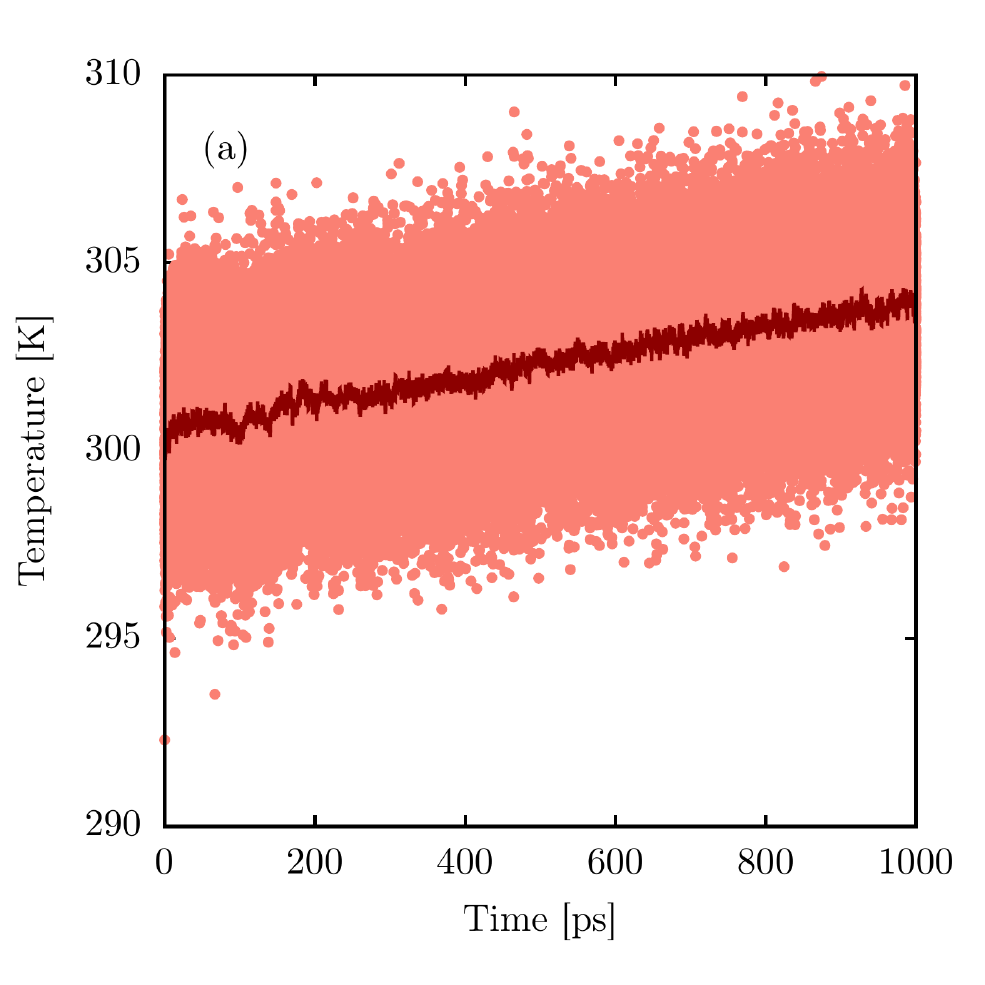}
\includegraphics[width=0.48\linewidth]{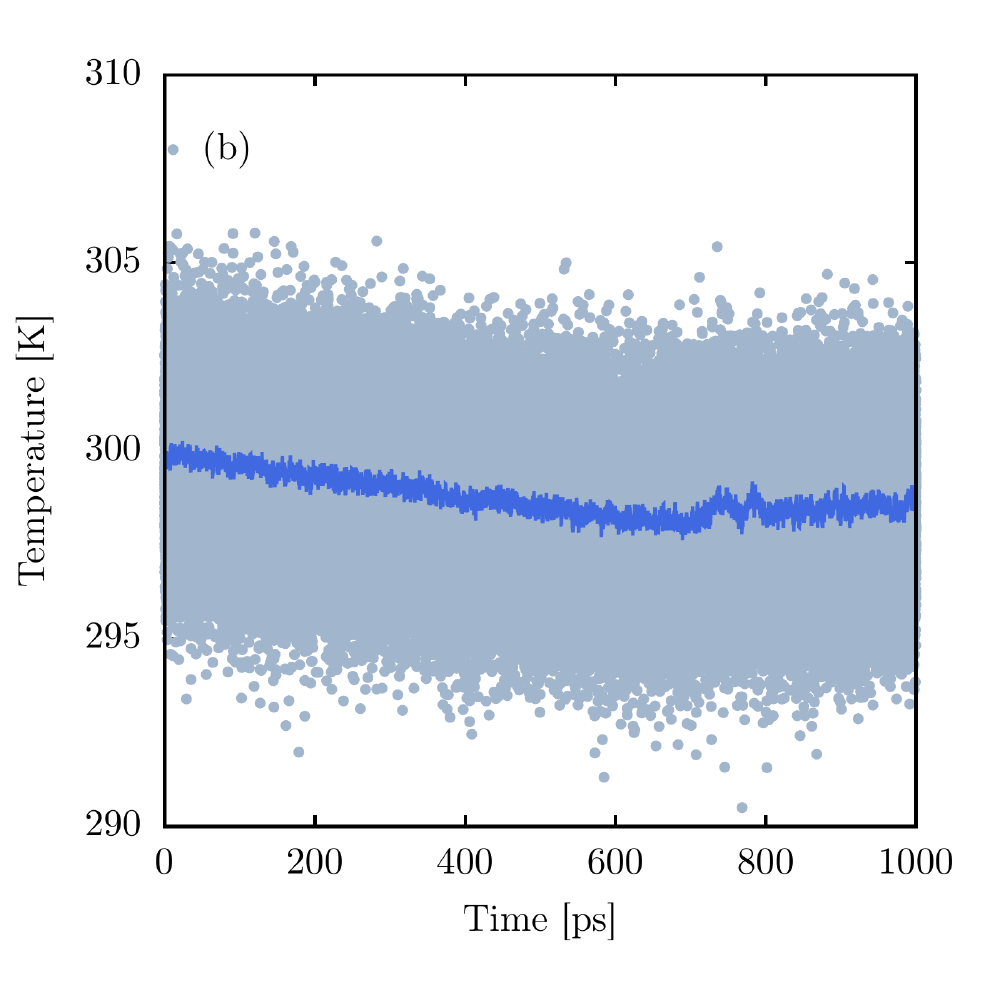}
\includegraphics[width=0.53\linewidth]{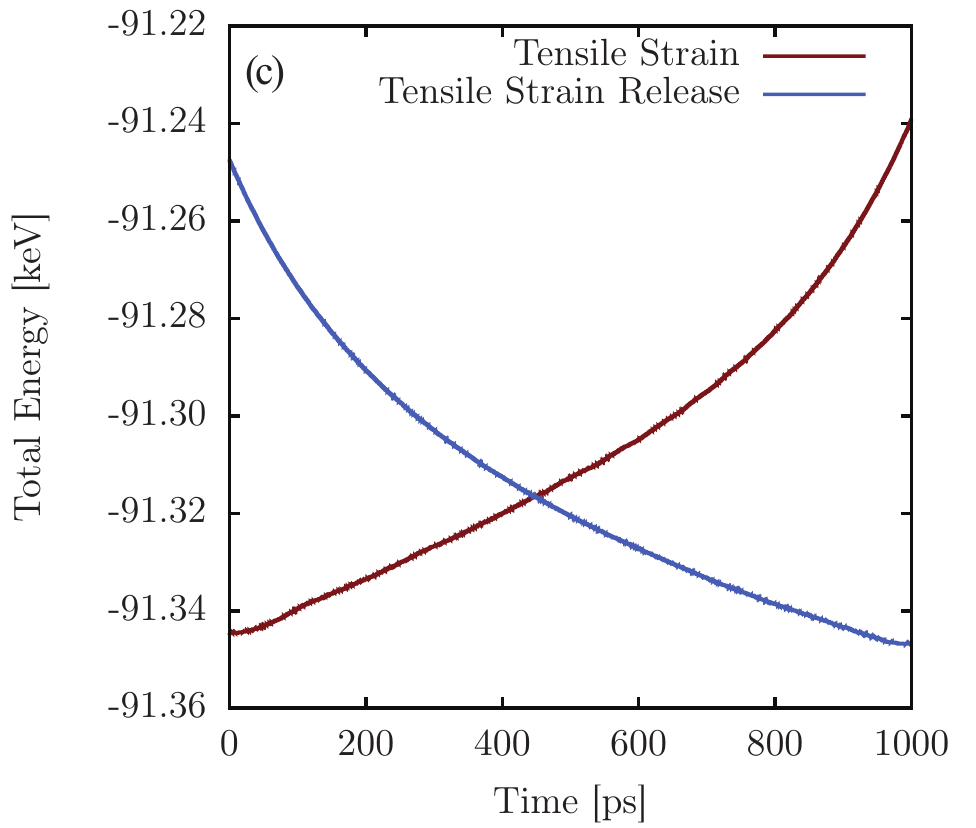}
\caption{Time evolution of the GK temperature under (a) tensile stretching and (b) releasing for one particular seed. The red and blue lines are the moving averages for temperature calculated every 100 MD steps. Panel (c) illustrates the time evolution of total energy as a function of time. The red and blue curves refer to adiabatic tensile strain and the tensile strain release, respectively. }
\label{fig4}
\end{center}
\end{figure}

We conducted ten simulations with different seeds for all the thermodynamic cycles, see Supplementary Material. For the sake of convenience and to present the qualitative behavior of the entire process, we show the time evolution of the GK temperature under tensile stretching and releasing for just one seed in Figures \ref{fig4}(a) and \ref{fig4}(b), respectively. This figure's red and blue lines represent the moving averages for temperature calculated every 100 MD steps. We obtained the moving averages for all seeds and related average values for $\Delta T_\text{MD}$ and used the $(c_\text{MD}/c_\text{REAL})=2.53$ factor to find $\Delta T_\text{REAL}\simeq 9.32$ K during the $1\rightarrow 2$ tensile stretching process, and $\Delta T_\text{REAL}\simeq -3.50$ K for the $3\rightarrow 4$ releasing process. The smoothed curves indicate that the GK temperature has an almost linear relationship with time. Additionally, it can be inferred from Figure \ref{fig4} that the ECE in GK is more significant during heating than during cooling. 

The GK coefficient-of-performance (COP) associated with heating or cooling can be calculated. To do that, we need to determine both the heat, $Q$, exchanged with a thermal bath and the work, $W$, %($W_\text{MD}$) 
done on and by the system during one full cycle. %, as well as its $C_\text{MD}$. 
Processes $2\rightarrow 3$ and $4\rightarrow 1$ are isochoric and do not contribute to the total work per cycle, while processes $1\rightarrow 2$ and $3\rightarrow 4$ are adiabatic, meaning there is no heat exchange into or out of the GK monolayer. Therefore, the total energy change due to tensile or tensile-release processes will equal the work done on or by the GK system. The total energy values as a function of simulation time during the $1\rightarrow 2$ tensile stretching process and the $3\rightarrow 4$ releasing process are shown in Figure \ref{fig4}(c), also just for one seed. $W$ can, then, be calculated from these curves as the sum of the variations in the internal energy of the GK during processes $1\rightarrow 2$ and $3\rightarrow 4$, denoted as $E_{1\rightarrow2}$ and $E_{3\rightarrow4}$, respectively. Specifically, their average values considering all the seeds are $W=E_{1\rightarrow2}-E_{3\rightarrow4}=109.5-(-103.1)=212.6$ eV. 

\begin{figure}[htb!]
\begin{center}
\includegraphics[width=0.9\linewidth]{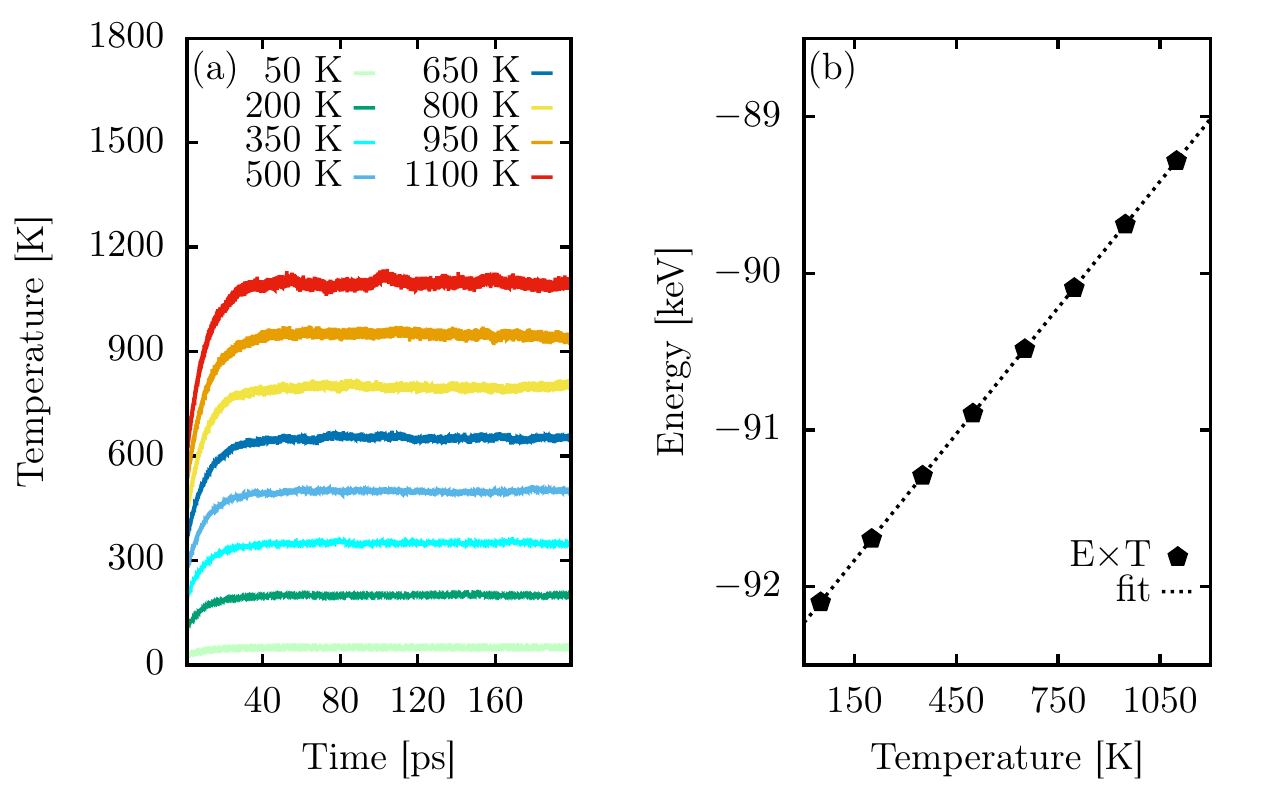}
\caption{(a) Time evolution of the temperature (GK thermalization) and (b) the total energy of the system as a function of its temperature. The temperature values used in these simulations vary from 50 K up to 1100 K with a step of 150 K.}
\label{fig5}
\end{center}
\end{figure}

To obtain the heat, $Q$, exchanged with the thermal bath, the thermal capacity, $C$, of the GK was estimated using a series of equilibrium simulations at different temperature regimes, as shown in Figure \ref{fig5}. The thermalization of the GK over time and the relationship between the temperature of the system and its total energy are shown in Figures \ref{fig5}(a) and \ref{fig5}(b), respectively. Temperature values ranged from 50 K to 1000 K in increments of 150 K. Linear fitting of the data in Figure \ref{fig5}(b) resulted in a thermal capacity of $C_\text{MD}=\Delta E/\Delta T=2.67$ eV/K, which divided by the mass of the GK and changing the units give the value previously mentioned of $c_\text{MD}=1770$ $\mathrm{J}\cdot\mathrm{Kg}^{-1}\cdot\mathrm{K}^{-1}$. With $c_\text{MD}$ and the values of $\Delta T_\text{MD}$, one can calculate the heat transferred/extracted ($Q_\text{T}$/$Q_\text{E}$) to/from the thermal reservoir at 300 K: $Q_\text{E}=mc_\text{MD}\Delta T_\text{MD}$ and $Q_\text{T}=mc_\text{MD}\Delta T_\text{MD}$.   

Finally, we calculate the COP for GK. Using the values derived above for $Q$ and $W$, we obtain the COP for heat transfer and extraction (COP$_\text{T}$/COP$_\text{E}$) to/from the thermal reservoir: COP$_\text{T}=Q_\text{T}/W_\text{MD}$ and COP$_\text{E}=Q_\text{E}/W_\text{MD}$. Some alloys, for instance, NiTi wires, show an ECE of about 17 K with a large COP of about 11~\cite{cui2012ECENiTi}. Carbon nanotubes can exhibit an ECE varying between 40 K to 100 K with also a large COP that ranges from 4 to 12~\cite{TatianaPRB2022}. GK has reasonable temperature change values of approximately 9.32 K for heating and -3.50 K for cooling, as mentioned above, accompanied by an equally reasonable COP of approximately 1.57 and 0.62, respectively. Although these values are not so impressive as that of NiTi wires or carbon nanotubes, they represent about 23 and 8.8 times larger ECE temperature changes than that of macroscopic kirigami materials~\cite{hiraiADVFMATT2022}. The Supplementary Material provides the values for the aforementioned quantities for individual seeds, which were used to determine the COP. 

%\textcolor{red}{COP is usually compared to the performance of an ideal Carnot refrigerator, COP$_\text{CARNOT}=T_\text{C}/(T_\text{H}-T_\text{C})$, operating between thermal reservoirs at cold (C) and hot (H) temperatures, $T_\text{C}$ and $T_\text{H}$, respectively. In this sense, for the GK, we obtained averaged values of $\textrm{COP}^\textrm{CARNOT}_\textrm{C} = 240.16$, $\textrm{COP}^\textrm{CARNOT}_\textrm{H} = 84.19$. 

\section{Conclusion}

In summary, our study investigated the ECE in a GK monolayer at the nanoscale, representing the thinnest possible KM. We compared the temperature changes of GK with macroscopic KMs, which was experimentally realized recently \cite{hiraiADVFMATT2022}, and identified key differences. Moreover, this study is the first to determine the COP in GK.

Through fully-atomistic molecular dynamics simulations, we found that GK exhibited a relatively large temperature change of approximately 9.32 K for heating and -3.50 K for cooling. This temperature change is significantly higher than the observed temperature changes in macroscopic KMs, which reached a maximum of approximately 0.4 K. The one-atom-thick configuration of graphene in GK prevented the complex temperature variations observed in macroscopic KMs, where simultaneous heating and cooling occurred in different regions when the external strain was applied.

Our simulations also revealed that the deformation mechanism of GK involved elastic stretching of C-C bonds and out-of-plane buckling. The out-of-plane deflected regions did not experience cooling or warming, contrary to the behavior observed in macroscopic KMs. This disparity arose because flexing a one-atom-thick nanoribbon did not generate strains in the C-C bonds, unlike macroscopic kirigami structures.

We calculated the COP of GK, representing the ratio of heat transferred to/from the thermal reservoir to work performed on the system per thermodynamic cycle. Our results demonstrated a reasonable COP of approximately 1.57 (0.62) for GK as a solid heater (refrigerator), indicating its potential for %efficient 
heat transfer and extraction. When comparing the COP of GK to that of other materials, such as alloys and carbon nanotubes, GK presents a smaller performance. However, neither alloys nor nanotubes reversibly withstand a tensile strain as high as 30\%. That feature might be considered when considering smart applications of GK as solid heaters or refrigerators at the nanoscale. Our findings highlight the unique behavior of GK at the nanoscale and its potential as a kirigami-based material for applications requiring substantial temperature changes and efficient heat transfer.

\section*{Acknowledgements}

This work received partial support from Brazilian agencies CAPES, CNPq, and FAPDF. L.A.R.J thanks the financial support from Brazilian Research Council FAP-DF grants $00193-00000857/2021-14$, $00193-00000853/2021-28$, and $00193-00000811/2021-97$, CNPq grants $302236/2018-0$ and $350176/2022-1$, and FAPDF-PRONEM grant $00193.00001247/2021-20$. L.A.R.J also thanks ABIN grant 08/2019 and Núcleo de Computação de Alto Desempenho (NACAD) for computational facilities through the Lobo Carneiro supercomputer. AFF is a fellow of the Brazilian Agency CNPq-Brazil ($303284/2021-8$) and acknowledges grant \#$2020/02044-9$ from São Paulo Research Foundation (FAPESP). This work used resources of the Centro Nacional de Processamento de Alto Desempenho em S\~ao Paulo (CENAPAD-SP). The authors acknowledge the National Laboratory for Scientific Computing (LNCC/MCTI, Brazil) for providing HPC resources of the SDumont supercomputer, contributing to the research results reported within this paper. URL: http://sdumont.lncc.br.

\bibliographystyle{unsrt}
\bibliography{bibliography.bib}
	
\end{document}

\title{Supplementary Information for\\Elastocaloric Effect in Graphene Kirigami}

\author{Luiz A. Ribeiro Junior}
 \affiliation{University of Bras\'{i}lia, Institute of Physics, Bras\'{i}lia, Brazil}
 
\author{Marcelo L. Pereira Junior}
 \affiliation{University of Bras\'{i}lia, Faculty of Technology, Department of Electrical Engineering, Bras\'{i}lia, Brazil.}
 
 \author{Alexandre F. Fonseca}
 \affiliation{Applied Physics Department, Gleb Wataghin Institute of Physics, University of Campinas, Campinas, S\~ao Paulo, Brazil.}

\maketitle

\begin{figure}
    \centering
    \includegraphics[width=0.8\linewidth]{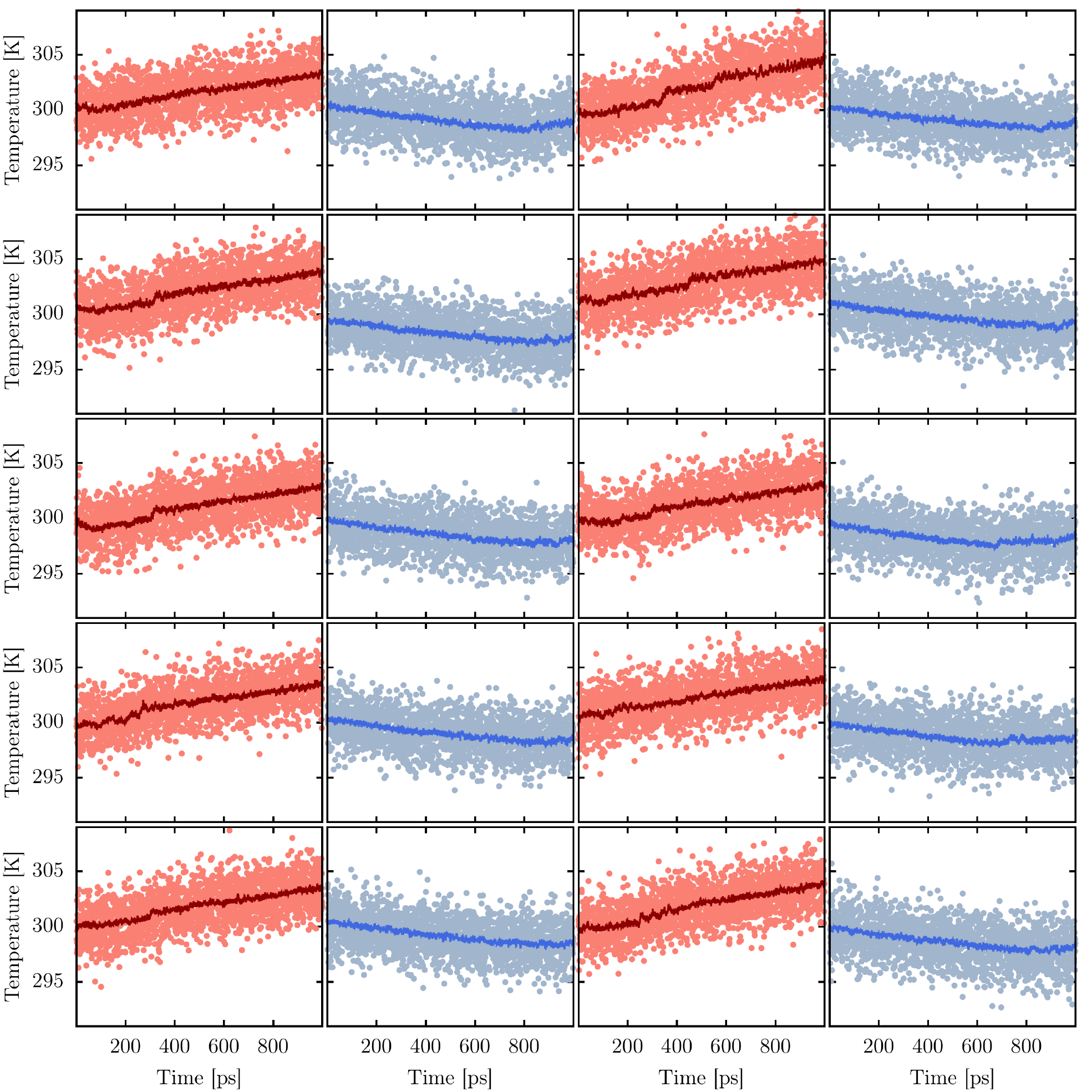}
    \caption{Time evolution of the GK temperature under (red) tensile stretching and (blue) releasing for ten different seeds. The red and blue lines are the moving averages for temperature calculated every 100 MD steps.}
    \label{fig:S1}
\end{figure}

\begin{table}[]
\begin{tabular}{|c|c|c|c|c|c|c|c|}
\hline
Simulation                       & \multicolumn{1}{c|}{$\Delta T_\textrm{MD}$} & \multicolumn{1}{c|}{$\Delta E_\textrm{MD}$} & \multicolumn{1}{c|}{$\textrm{Q}_\textrm{f}$} & \multicolumn{1}{c|}{$\textrm{W}_\textrm{c}$} & \multicolumn{1}{c|}{COP} & COP$_\textrm{CARNOT}$ & \multicolumn{1}{c|}{$\Delta T_\textrm{REAL}$} \\ \hline
\multirow{10}{*}{Strain}         & 2.81                                       & 112.93                                     & 7.56                                         & 5.27                                         & 1.43                     & 107.64                & 7.12                                         \\ \cline{2-8} 
                                 & 4.94                                       & 99.09                                      & 13.28                                        & 10.62                                        & 1.25                     & 61.69                 & 12.51                                        \\ \cline{2-8} 
                                 & 3.66                                       & 110.06                                     & 9.84                                         & 5.55                                         & 1.77                     & 82.87                 & 9.27                                         \\ \cline{2-8} 
                                 & 3.43                                       & 105.53                                     & 9.21                                         & 6.23                                         & 1.48                     & 88.50                 & 8.67                                         \\ \cline{2-8} 
                                 & 3.24                                       & 109.80                                     & 8.70                                         & 5.97                                         & 1.46                     & 93.60                 & 8.20                                         \\ \cline{2-8} 
                                 & 3.23                                       & 109.43                                     & 8.66                                         & 7.05                                         & 1.23                     & 94.02                 & 8.16                                         \\ \cline{2-8} 
                                 & 3.97                                       & 104.85                                     & 10.66                                        & 5.60                                         & 1.90                     & 76.62                 & 10.04                                        \\ \cline{2-8} 
                                 & 3.78                                       & 109.53                                     & 10.15                                        & 6.40                                         & 1.59                     & 80.35                 & 9.57                                         \\ \cline{2-8} 
                                 & 3.83                                       & 109.47                                     & 10.28                                        & 5.00                                         & 2.05                     & 79.42                 & 9.68                                         \\ \cline{2-8} 
                                 & 3.94                                       & 105.54                                     & 10.58                                        & 6.91                                         & 1.53                     & 77.18                 & 9.96                                         \\ \hline
\multirow{10}{*}{Strain Release} & -1.55                                      & -107.65                                    & -4.17                                        & 5.27                                         & 0.79                     & 192.13                & -3.93                                        \\ \cline{2-8} 
                                 & -1.19                                      & -88.47                                     & -3.19                                        & 10.62                                        & 0.30                     & 251.84                & -3.00                                        \\ \cline{2-8} 
                                 & -1.40                                      & -104.51                                    & -3.76                                        & 5.55                                         & 0.68                     & 213.28                & -3.54                                        \\ \cline{2-8} 
                                 & -1.42                                      & -99.29                                     & -3.82                                        & 6.23                                         & 0.61                     & 209.76                & -3.60                                        \\ \cline{2-8} 
                                 & -1.96                                      & -103.83                                    & -5.26                                        & 5.97                                         & 0.88                     & 152.13                & -4.96                                        \\ \cline{2-8} 
                                 & -0.58                                      & -102.38                                    & -1.56                                        & 7.05                                         & 0.22                     & 516.09                & -1.47                                        \\ \cline{2-8} 
                                 & -1.73                                      & -99.25                                     & -4.64                                        & 5.60                                         & 0.83                     & 172.75                & -4.37                                        \\ \cline{2-8} 
                                 & -1.06                                      & -103.13                                    & -2.85                                        & 6.40                                         & 0.45                     & 281.59                & -2.69                                        \\ \cline{2-8} 
                                 & -1.69                                      & -104.46                                    & -4.54                                        & 5.00                                         & 0.91                     & 176.33                & -4.28                                        \\ \cline{2-8} 
                                 & -1.27                                      & -98.63                                     & -3.40                                        & 6.91                                         & 0.49                     & 235.65                & -3.21                                        \\ \hline
\end{tabular}

\label{tabs1}
\caption{Quantities used to determine the COP for each seed value: simulated temperature change ($\Delta T_{MD}$), total energy variation in the MD simulation ($\Delta E_{MD}$), transferred heart ($Q_f$), work done on and by the system during one full cycle ($W_C$), coefficient of performance (COP), and coefficient of performance of an ideal Carnot refrigerator ($COP_\text{CARNOT}$). The ECE temperature variation ($\Delta T_\text{REAL}$) from classical MD simulations is corrected due to quantum mechanics effects on the specific heat ($C$) of graphene at 300 K \cite{LisenkovNANOLETT2016}: $\Delta T_\text{REAL}=(C_\text{MD}/C_\text{REAL})\Delta T_\text{MD}$. Here, ``REAL" and ``MD" refer to corrected and simulated values, respectively. The 2.53 factor was obtained using $C_\text{REAL}=700$ $\mathrm{J}\cdot\mathrm{Kg}^{-1}\cdot\mathrm{K}^{-1}$ from Ref.~\cite{liNANOSCALE2017} and $C_\text{MD}=1770$ $\mathrm{J}\cdot\mathrm{Kg}^{-1}\cdot\mathrm{K}^{-1}$ from our calculations. $C_\text{MD}$ was derived from a series of equilibrium simulations for GK subjected to distinct temperature regimes, as discussed later.} 
\end{table}